# Configuration management in the distributed cloud

Tamara Ranković[0000-0002-5265-4626], Ivana Kovačević[0000-0001-9418-8814], Veljko Maksimović[0000-0002-4905-2421], Goran Sladić[0000-0002-0691-7392] and Miloš Simić[0000-0001-8646-1569]

Faculty of Technical Sciences, University of Novi Sad, Serbia
`tamara.rankovic@uns.ac.rs`

**Abstract.** Owing to their cost-effectiveness and flexibility, cloud services have been the default choice for the deployment of innumerable software systems over the years. However, novel paradigms are beginning to emerge, as the cloud can't meet the requirements of increasingly many latency- and privacy-sensitive applications. The distributed cloud model, being one of the attempts to overcome these challenges, places a distributed cloud layer between device and cloud layers, intending to bring resources closer to data sources. As application code should be kept separate from its configuration, especially in highly dynamic cloud environments, there is a need to incorporate configuration primitives in future distributed cloud platforms. In this paper, we present the design and implementation of a configuration management subsystem for an open-source distributed cloud platform. Our solution spreads across the cloud and distributed cloud layers and supports configuration versioning, selective dissemination to nodes in the distributed cloud layer, and logical isolation via namespaces. Our work serves as a demonstration of the feasibility and usability of the new cloud-extending models and provides valuable insight into one of the possible implementations.

## 1 Introduction

A paradigm shift from on-premise to cloud deployments revolutionized the software industry. Capital expenditures plummeted, which allowed many to host their solutions without the need for ahead-of-time costly investments in infrastructure. In the cloud, compute, storage, and network resources are rented on demand and released when unnecessary, with the user being charged only for what they'd used. Such a payment model yielded financial benefits both for cloud providers and their clients. To ensure profitability, providers keep all infrastructure in a few large data centers, placed in locations strategically selected to lower the management costs [1].

In recent years, use cases showcasing the limitations of the cloud's centralized resource organization emerged. For the increasing number of systems relying on real-time data processing, such as industrial Internet of Things (IoT), IoT in healthcare, and autonomous vehicles, it has been repeatedly demonstrated that the latency of sending data to and processing it in the cloud is too high [2]. Moreover, many applications handling sensitive user data have to comply with rigorous legal regulations regarding data placement [2], which can be challenging considering how sparsely spread data centers can be.

This pushed the research community and industry to put forward solutions that eliminate, or at least alleviate the mentioned difficulties. The dominant line of thought is that resources should become more geographically distributed [3]. This new organization doesn't make the cloud obsolete, rather it compliments it by providing a layered model that satisfies latency and data privacy requirements, but also offers all benefits of the traditional cloud. Edge [4] and fog computing [5] surfaced as such models, putting a layer at the edge of the network, between the cloud and clients, to bring compute and storage closer to data sources.

Another model, motivated by the same problems, is the distributed cloud (DC) model [6, 7]. It expands the cloud model by placing a distributed cloud layer between the cloud and device layers. In that sense, it resembles the edge and fog models. However, the distributed cloud layer has a clearly defined hierarchical structure. The smallest unit of resources is a cluster. Multiple clusters comprise a region - a logical collection of clusters. Likewise, multiple regions comprise a topology. The three-level organization contributes to the robustness of the system. With this model, users can form ad-hoc clouds consisting of nodes selected to meet their specific needs. They can be utilized on their own or in conjunction with the cloud. This opens the door for many opportunities but requires a substantial effort when it comes to the implementation of a platform that would follow this model.

The focus of this paper will be the design and implementation of the module responsible for configuration management in the DC environment. The module is embedded in the Constellations (c12s) open-source DC platform.1 When we say configuration, we refer to the specification of application parameter values that determine its runtime behavior. There are multiple strategies for providing configuration to the application, the simplest one being hard coding. This approach leads to low portability and is considered an anti-pattern, as any configuration update requires source code modifications. Alternative strategies include configuration files and injection via environment variables, with the latter being recommended by *The Twelve-Factor App* guidelines for building cloud-native applications [8]. A large body of papers concerned with configuration errors witnesses the importance of a streamlined configuration process that can prevent and detect as many foreseeable errors as possible, especially in environments as complex as the DC is [9].

The rest of the paper is organized as follows: Section 2 discusses how configuration management is handled in environments similar to DCs, while Section 3 delves into details of how it is implemented in the DC. Section 4 presents limitations of the current solution and directions of future work. In the end, Section 5 concludes the paper.

---

[1] https://github.com/c12s

## 2   Related work

In this section, we'll give an overview of configuration management solutions for different systems and environments. Based on these, we extracted requirements for the DC platform and detected common challenges that need to be addressed.

A comprehensive case study by the Facebook team [10] provides insight into the design and implementation of their configuration management system, handling thousands of changes and trillions of checks every day. The Configurer centralized service is a core component responsible for version control, schema validation, and dissemination. All configuration is stored in Git. The git repository is monitored for changes that, when detected, are written to a variant of ZooKeeper [11]. Each server runs a Proxy process, communicating with applications on that server. On startup, applications ask the Proxy for configuration, which then fetches it and watches for future updates. The authors opted for the push model for scalability reasons. As it is challenging to ensure scalable delivery for large configuration files in this way, they adhere to an alternative distribution strategy. For files larger than 1MB, the exchange occurs in a peer-to-peer manner, relying on the BitTorrent protocol [12].

This system takes the *configuration-as-code* approach. The engineers specify configuration as Python code that later gets compiled by the Configurer into a JSON file. The authors of [13], while reflecting on the development of the Borg cluster manager [14], also deemed this approach to be unavoidable in systems with versatile requirements. They advise embracing it when necessary and relying on general-purpose languages for programmatic manipulations, as they offer richer ecosystems than the domain-specific ones, designed only for configuration management.

Kubernetes [15], the de facto dominant open-source container orchestration tool right now, also follows the principle of separating code from configuration and uses the ConfigMap primitive to achieve so. Users first specify the ConfigMap resource, and later bind it to a pod by referencing it in its specification. ConfigMap values are supplied to applications running inside a pod in one of two ways: 1) Via environment variables 2) By mounting a volume in which each key-value pair is represented as a file. When values are injected through environment variables, each ConfigMap change requires a pod restart. On the other hand, a volume will periodically get updated with new values, which allows for dynamic reconfiguration of the application. However, no Kubernetes component will notify the application of the change, so a pull model has to be employed.

Akamai is a content delivery network (CDN) with over 15,000 servers hosting web content and applications [16]. Akamai clients are able to configure many parameters, such as HTML cache timeouts and whether to allow cookies or store session data. The Akamai Configuration Management System (ACMS) [16] was developed to facilitate fault-tolerant, consistent updates with efficient and secure delivery. The system consists of *publishers*, who can concurrently push new versions of configuration files, *storage points*, that receive, accept, persist, and replicate configuration, and *receivers*, that run on each node and fetch configuration from storage points. Applications running on nodes connect to their receiver and subscribe to a certain configuration file. After that, the receiver periodically pulls new versions from storage points. The pull-based approach was selected because the Akamai CDN is fully optimized for HTTP download and caching capabilities can be utilized to reduce network bandwidth requirements.

## 3 Configuration management

The main objective of the platform is to enable cloud-like services to users who would benefit from highly elastic as well as latency- or privacy-aware deployments. To achieve so, it should offer streamlined processes for infrastructure provisioning, application life cycle, and behavior management. Such a task requires proper handling of numerous resource types in a multi-tenant and geographically distributed environment.

Section 3.1 introduces and describes the most significant properties of system resources relevant to this paper, whereas in Section 3.2 we discuss the architecture of the platform and how node and configuration management operations have been executed.

### 3.1 Resources

Typical clients interacting with cloud platforms are developers, DevOps, Platform or Site Reliability Engineers (SREs), all being **users** with different responsibilities. They usually collaborate in groups, on one or multiple projects, so infrastructure provisioned by one platform engineer should be available for others to test or deploy their applications on, for example. To facilitate this workflow, through the **organization** resource we keep track of all users affiliated with that organization. Then, infrastructure provisioned by a member of the organization, and all resources subsequently associated with that infrastructure become ownership of the organization, so other members can leverage it. Also, security policies are administered on the organization level, so no unprivileged operations are to take place.

We regard infrastructure as a set of **nodes**, each equipped with a certain degree of storage, computing, and networking capability. A node can be in one of two states:

- **Available** - Currently in the *node pool*, waiting to be taken by some organization.
- **Occupied** - In the ownership of an organization. If the node is occupied, it is so by exactly one organization.

Every node joins the system with a predefined set of labels, encoding the node's hardware, software, and location information. Structurally, a label is just a key-value pair, but semantically it serves to describe a node more precisely. Predefined labels are read only from the client's perspective, but the label set can be enriched with additional user-specified labels. Those labels should hold attributes of the node meaningful to its operators and defined to facilitate the management process. It should be noted that a user can set a label only if the target node is owned by their organization and if they were granted the necessary privileges.

Labels provide a powerful mechanism for querying nodes for different purposes, such as claiming them or disseminating configuration. A *query* is a set of *label selectors*, each consisting of a logical expression. When a selector evaluates to *true* for a label, we say the label matches the selector. When applied to a set of nodes, a label selector generates a new set, containing nodes from the original set for which the logical expression is *true*. For a query that comprises multiple label selectors, the final result set is obtained as the intersection of each selector's result set. This equates to the AND operation. If the behavior of the OR operation is desired, it can be accomplished by running multiple queries and combining the results. A more formal query definition is provided below:

$$query = \{label\_selector \mid label\_selector \in label\_keys \times op \times label\_values\}$$

$$op = \{=, <, >, \leq, \geq\}$$

When an organization acquires infrastructure, its members can run workloads or store data on it. In this paragraph, we will only briefly introduce the concept of **applications** and **namespaces** to the reader, as they are going to be an integral part of the platform. However, they are subjects of ongoing research and only broad details are necessary for understanding the remainder of this paper. Applications are executable programs that can run in a specified isolated environment, such as containers. Aside from that condition, there are no restrictions on the technology stack the application must or mustn't rely on. From the perspective of one node, the platform currently doesn't allow interorganizational multitenancy, but there is still intraorganizational multitenancy. Namespaces aim to provide control over such an environment through logical isolation of applications and data. Every application, data unit, as well as configuration, belongs to a namespace. By default, elements from one namespace are invisible and unreachable to elements of the other namespaces. Even inside one namespace, security policy administrators explicitly specify what privileges applications have.

The central resource to this paper is **configuration** - an argument that dictates the runtime behavior of a parametrized application that consumes it. The format of a configuration is quite simple, it is a key-value pair with both elements represented as text. Users are free to choose any data type for the configuration value, but they are obliged to serialize it to a text format when submitting a new configuration. Likewise, applications consuming it need to have deserialization logic implemented. A configuration is not declared individually, but as part of a *configuration group*, which is a struct consisting of:

- **Name** - A unique identifier for the group inside an organization.
- **Organization** - The owner of the configuration group. A configuration group can be consumed only by applications from the same organization.
- **Configurations** - A set of key-value pairs.
- **Version** - A version of the configuration group. Whenever a configuration inside the group is to be added, updated, or removed, a new version must be set. This makes tracking changes over time more straightforward. Also, different versions of an application may need to consume different versions of the same configuration group.

Configuration groups are initially only stored in the control plane. They have to be applied to be disseminated to nodes. Through configuration dissemination, the users declaratively specify to what nodes a configuration group version will be propagated and in what namespace it will be visible. A detailed explanation of this process will be given in the next section.

### 3.2 System overview

In this section, we'll discuss how operations regarding configuration management were implemented. Firstly, we'll present the list of software components the platform consists of. After that, we'll go into details of request flows of node registration and ownership claiming, as their execution sets up the environment for successful configuration dissemination, which we'll present in the end.

The platform can be divided into two distinct units, each with their responsibilities:

- **Control plane** - The brain of the system that receives and handles all end-user requests. In doing so, it issues commands to the nodes, so that the state of the infrastructure converges towards the desired state, specified by the users. The control plane should be deployed in an environment abundant in resources, such as the cloud.
- **Node agents** - Lightweight processes running on each node, responsible for establishing a connection to the control plane, receiving commands from it, acting upon those commands, and reporting on the outcome. In this paper, we'll focus on bootstrap and configuration agents.

The control plane is implemented modularly, as a collection of microservices written in Go, and with system extendability in mind. In **Error! Reference source not found.**, we can see what services the current proof-of-concept implementation contains and what their responsibilities are. Services communicate with each other synchronously, via gRPC[2], or asynchronously, relying on the NATS[3] message broker for delivery. The clients are provided with a REST API, with all requests being translated into gRPC by the API gateway. The node agents were also implemented in the Go language and are currently packaged as a single binary per node. Currently, communication between the control plane and node agents is facilitated by NATS, which can be swapped for any other transport solution guaranteeing message delivery.

Before any operations invoked by the user take place, the system has to ensure that all nodes are visible to the control plane, establish communication channels between them, etc. We refer to this working mode of the system as a *setup mode*, while interactions with users are carried out in the *operational mode*. As processes in the setup mode should precede the operational mode, we'll first delve into implementation details of the former, and subsequently the latter.

---

[2] gRPC (https://grpc.io/)
[3] NATS.io – Cloud Native, Open Source, High-performance Messaging (https://nats.io/)

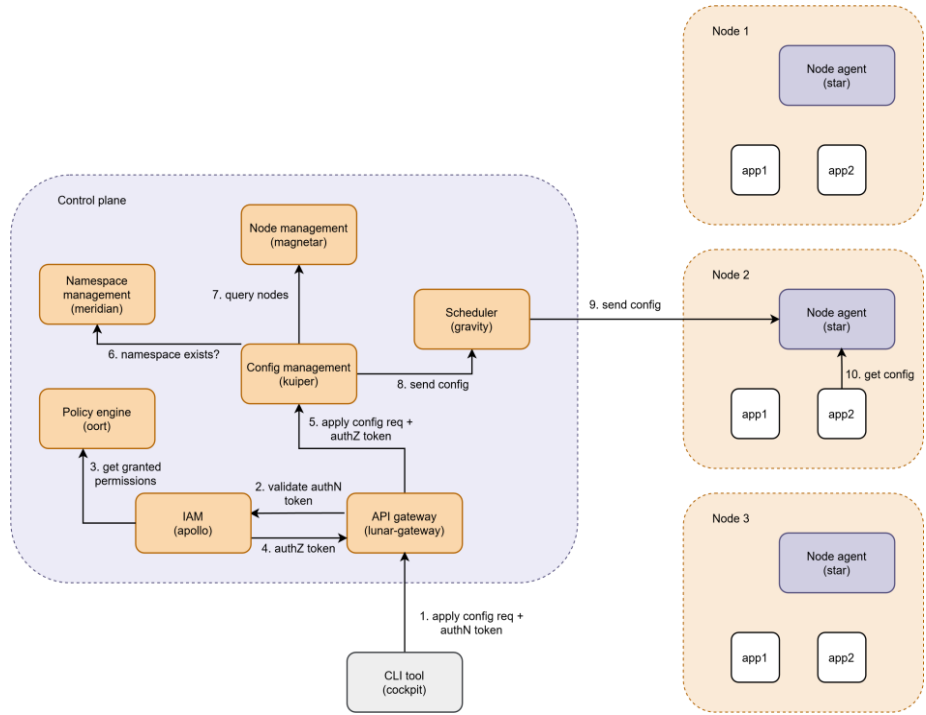

**Fig. 1.** System architecture with a request flow of the apply configuration operation

**Setup mode.** For the control plane to be able to manage nodes' state, it needs to be aware of their presence in the node pool. Hence, a node discovery mechanism is required. Each node undergoes a registration process invoked by a bootstrap node agent. On startup, the agent checks if the node has previously been registered. It does so by searching for a *nodeid* file in the */etc/c12s* directory. The file should contain a unique identifier of the node assigned upon registration. If it finds it successfully, there is no need for any further action, as the node has already introduced itself, but encountered a crash or intentional restart sometime after that. If, however, the identifier couldn't be found, the bootstrap agent prepares and sends a registration request. In the preparation phase, all relevant data regarding the node's hardware, software, and location are collected. Based on this, the initial label set is formed. Then, a request is sent to the *registration* NATS subject, for the control plane's *magnetar* service to receive it. The recipient service generates a unique identifier for the node, marks it as *available*, persists that together with the node's labels to the etcd[4] store, and publishes a response containing the identifier. The reply subject is specified in the request's metadata. After the bootstrap agent receives a response, it persists the identifier to the filesystem and notifies the configuration agent of the changes.

---

[4] https://etcd.io/

There is a communication channel between the control plane and each node for propagating the configuration. It is implemented as a NATS subject, named *config.{nodeid}*. This means the node has no way of receiving configuration before registration, which is sensible as only after that it becomes visible in the node pool.

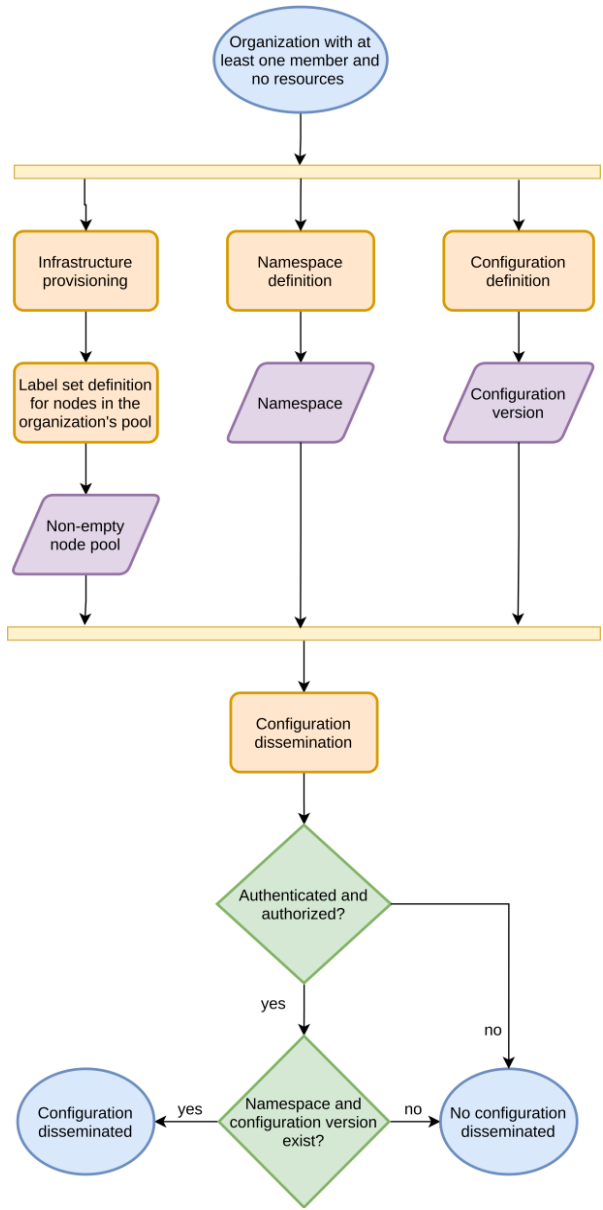

**Fig. 2.** Flowchart of the operational mode

All configuration-related tasks on nodes are handled by the configuration agent. It has two main responsibilities: 1. Receive new configuration group versions from the control plane 2. Serve requests from applications running on the node, trying to fetch a configuration group version. To function properly, the configuration agent needs to know the node identifier. So, when started, it first goes through an initialization process. It starts by searching for the *nodeid* file. If found successfully, the agent subscribes to the *config.{nodeid}* subject and starts a gRPC server for communication with applications running on the node. In the case when the identifier couldn't be found, the configuration agent halts until it receives a signal from the bootstrap agent that the identifier has been successfully set. Then, it retries the initialization process.

**Operational mode.** When users become affiliated with an organization, they can provision infrastructure for that organization and manage applications, configuration, and namespaces on that infrastructure. For any configuration to be disseminated to any node, several operations have to occur first. The organization needs to have nodes in their pool and at least one version of one configuration group specified. Additionally, other optional operations can be performed, such as application deployments and the creation of new labels or namespaces. We'll briefly discuss the mechanics of the mandatory ones to be executed.

In the provisioning process, nodes are transferred from the pool of available nodes to the pool of the organization. Nodes to be moved are determined by a label-based query, specified by the user in charge of provisioning. Matched nodes' statuses are changed from *available* to *occupied* and an ownership relationship between the organization and the node is established. The *magnetar* service implements this logic.

When creating a new configuration group version, a user must set all properties stated in Section 3.1. The operation is deemed successful only if the version is valid and the user has the necessary privileges to add a configuration group to the specified organization. A version is valid if there is no other configuration group with the same name and of the same version in the organization. Configuration versioning and persistence are handled by the *kuiper* service.

After the successful provisioning and configuration group specification, the user can disseminate a configuration group version to a subset of nodes from the node pool. Each request issued must contain the name of the organization for which the dissemination is to take place, the name and version of the configuration group, the namespace in which it'll be logically visible, and a query for selecting the nodes to propagate configuration to. The *apply configuration* operation is orchestrated by the *kuiper* service and requires interaction between multiple services. In Fig. 1, a successful request execution flow is given, while the entire workflow of the operational mode is displayed in Fig. 2. The steps involved in the dissemination request are the following:

- The request is intercepted by the API gateway which firstly validates the authentication token with the *apollo* service. If the token is valid, *apollo* will generate an authorization token, containing all current permission for the user, determined by the *oort* policy engine.

- Next, the API gateway will make an *ApplyConfig* RPC to *kuiper* and attach the authorization token to the request's metadata.
- *kuiper* will first check if the specified namespace exists. If not, the request will be aborted.
- After that, it will get nodes matching the query from *magnetar*.
- It will schedule the propagation of data to nodes with the *gravity* service
- *gravity* will publish a message containing configuration to the *config.{nodeid}* subject, with identifiers previously obtained from *magnetar*
- The configuration agent of each node will store configuration data extracted from the received message

After the configuration is successfully disseminated, applications running on those nodes can pull it by sending a gRPC request to the configuration agent.

## 4      Limitations and future work

Currently, the user can list configuration groups by their name, which effectively displays configurations by version. However, there is no clear view of the progression of versions through time. In the future, we want to provide users with a diff tool that can detect what changes were made in one version, relative to the previous version.

Because a single configuration is treated as an arbitrary key-value pair, the user can specify any values when submitting a new version. This can lead to misconfiguration for different reasons. For example, there may be agreed-upon constraints on the value type or range, which can easily be violated with no validation on the system side. Also, spelling errors or the use of lower-case instead of upper-case letters and vice versa can lead to the application not being able to find the necessary configuration. For those reasons, we plan to introduce schema specification and validation features so that the reconfiguration becomes less error-prone in the future. Each configuration group would have to conform to a certain schema, or else it would be rejected.

When a configuration group reaches a certain node, the application running on it is not aware of the change unless it queries the configuration agent. This pull-based system makes the agent more straightforward to implement but puts an additional burden on the developers of the applications. If the configuration is not stored on the node when the application starts, it has to ask for it periodically. This also introduces unnecessary chatter. Because of this, a push-based mechanism will be introduced in the future, so that applications can subscribe to required configuration group versions or version ranges, and the configuration agent will be responsible for delivering it to them.

As of now, the organization's infrastructure is one node pool shared by all of its members and workloads, and with no communication between the nodes. To utilize the cloud setting fully, one direction of future work is to support clustering. This will impact configuration dissemination as it will be necessary to deliver configuration to a cluster, not just a set of distinct nodes.

This will be executable using one of two possible strategies:

- Direct dissemination: The control plane will be responsible for delivering new configuration to all members of the specified cluster. This resembles the current dissemination process but with different criteria for selecting nodes.
- Dissemination via gossiping: The control plane will ensure delivery to a configurable percentage of cluster members. After that, it will delegate further dissemination to the cluster, relying on it to propagate data via a gossip protocol. This will offload the control plane to an extent and reduce the single-point-of-failure risk in certain cases.

Access control mechanisms are employed to ensure data confidentiality, but if unauthorized access is successfully performed, the attacker can obtain sensitive information such as passwords, API keys etc, as they are stored in plain text. To minimize the repercussions of potential authorization bypasses, we plan to extend configuration management with encryption capabilities.

## 5 Conclusion

In this paper, we tackled the problem of implementing a platform that allows users to form DCs, run workloads, and store data in them. More precisely, we dealt with the configuration management module, which has to ensure proper execution of configuration specification and dissemination to DCs. Our solution consists of a centralized configuration service in the cloud and agents running on each node. The user specifies what configuration should be present on what nodes, after which the platform makes sure the system state eventually reaches the desired state.

With cloud-edge models and platforms becoming more widespread, this paper contributes to their further research and development as it provides insight into mechanisms behind such solutions and serves as a demonstration of their feasibility and usability. Our future work is going to be directed towards enhancing the configuration management process and developing additional platform capabilities.

**Acknowledgements.**  Funded by the European Union (TaRDIS, 101093006). Views and opinions expressed are however those of the author(s) only and do not necessarily reflect those of the European Union. Neither the European Union nor the granting authority can be held responsible for them.